\documentclass[%
 reprint,
superscriptaddress,
 amsmath,amssymb,
 aps,
prb
]{revtex4-1}
\usepackage{graphicx,hyperref,amsthm,xcolor,soul,braket,todonotes,amssymb,amsmath}
\usepackage[]{subcaption}
\usepackage{dcolumn}
\usepackage{bm}
\usepackage{natbib}
\usepackage{ulem}

\pdfoutput=1
\setcitestyle{numbers,square}
\hypersetup{
    linkcolor=blue,
    citecolor=blue,
    filecolor=blue,
    urlcolor=blue,
    colorlinks=true
}
\DeclareMathOperator{\Tr}{Tr}

\newenvironment{nalign}{
    \begin{equation}
    \begin{aligned}
}{
    \end{aligned}
    \end{equation}
    \ignorespacesafterend
}

\begin{document}
	\title{Quantifying Unitary Flow Efficiency and Entanglement for Many-Body Localization}
	
	\author{Gregory A. Hamilton }
	\email{gah4@illinois.edu}
	\author{Bryan K. Clark}
	\affiliation{ 
Institute for Condensed Matter Theory and IQUIST and Department of Physics, University of Illinois at Urbana-Champaign, Urbana, IL 61801, USA}
	
	\date{\today}

	\begin{abstract}
We probe the bulk geometry of the Wegner Wilson Flow (WWF) in the context of many-body localization, by addressing efficiency and bulk entanglement growth measures through approximating upper bounds on the boundary entanglement entropy. We connect these upper bounds to the Fubini-Study metric and clarify how a central quantity, the information fluctuation complexity, distinguishes bulk unitary rotation from entanglement production. We also give a short new proof of the small incremental entangling theorem in the absence of ancillas, achieving a dimension-independent, universal factor of $c=2$.
 	\end{abstract}
\maketitle

\section{Introduction}\label{intro}

Many-body localization (MBL) is centrally a failure to thermalize \cite{pal2010,abanin2019colloquium,oganesyan2007localization}. At a phenomenological level, MBL is described by an emergent, macroscopic set of quasilocal integrals of motion, known colloquially as $\ell$-bits \cite{abanin2019colloquium,serbyn2013local,huse2014phenomenology}. These $\ell$-bits give rise to a myriad of effects, including vanishing conductivity, logarithmic entanglement growth, and area-law eigenstates \cite{pal2010,kim2014local,bardarson2012unbounded}. Constructing $\ell$-bits from local operators via a diagonalizing unitary $U$ establishes the stability of the MBL phase and yields insight on the dynamics of local observables \cite{imbrie2016many,de_Roeck_Imbrie_2017,roy2020fock,pekker2017encoding,yu2016bimodal,luitz2016long,yu2017finding,oganesyan2007localization,vznidarivc2008many,vznidarivc2016diffusive}. \par 

Considerable effort has been devoted in recent years towards finding appropriate $\ell$-bit construction methodologies, including strong-disorder renormalization flows, perturbative constructions, and infinite-time averaged local observables \cite{abanin2019colloquium,kim2014local,imbrie2016many,pekker2017fixed}. One recent approach, the Wegner Wilson flow (WWF), leverages a continuous renormalization flow that generally produces more quasilocal $\ell$-bits \cite{PhysRevB.100.115136,pekker2017fixed}. The breakdown of ergodicity upon entering the area-law, MBL phase implies an efficient tensor network construction of both eigenstates and $\ell$-bits \cite{chandran2015spectral,wahl2017efficient}. \par
The WWF renormalization flows boundary Hamiltonians to diagonalized form through a bulk unitary. Recent work \cite{yu2019bulk} considered aspects of this WWF flow implemented as a bulk unitary tensor network. The idea that tensor networks may define a quantum-informational bulk geometry is particularly relevant with respect to the Ryu-Takayanagi (RT) conjecture, frequently considered in the context of holographic tensor networks and continuous multiscale renormalization ansatzes (cMERAs) \cite{pastawski2015holographic,miyaji2015surface,qi2013exact,swingle2012entanglement,witten1998anti,maldacena1999large}. However, this connection between bulk geometry and boundary entanglement is less clear in a generic tensor network construction. Beyond the tensor network description, the WWF has strong ties to quantum geometry and geodesicity in the projective Hilbert space \cite{itto2012geodesic}.

In this work, we examine bulk measures derived from the WWF, with a focus on the many-body localized phase of matter. We begin in Sec. \ref{sec2} by examining not a bulk geometry, but rather a speed limit for the entanglement entropy. We use the small incremental entangling (SIE) theorem as a starting point to probe the entanglement structure and efficiency of unitary flows. Along the way, we give a simplified proof of the ancilla-free version of the SIE, and achieve a tighter bound than the state of the art. Inspired by the RT conjecture, which connects the boundary entanglement with distances in the holographic bulk, we demonstrate how the SIE theorem identifies computable bulk metrics intimately tied to notions of efficiency and quantum complexity. In  Sec. \ref{sec3}, we apply the SIE theorem and our bulk metrics to the WWF in the context of an MBL system. We show that the WWF is monotonically more efficient upon entering the MBL phase, and that the boundary entanglement entropy monotonically decreases with efficiency. What is more, we show that some bulk metrics, while easily computable, fail to fully diagnose the flow entanglement dynamics. We identify a key quantity, the information fluctuation complexity, as a source of this failure.

\begin{figure}
    \centering
    \includegraphics[width = \linewidth,trim = 1.0cm 0cm .8cm 1cm,clip]{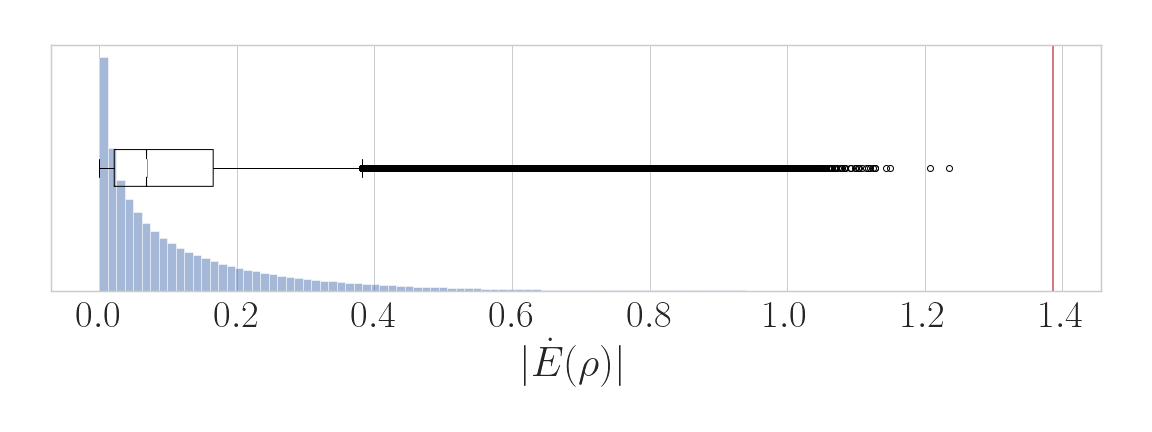}
    \caption{Histogram and box plot of $\dot{E}(\rho)$ for a two-qubit system, over two million realizations of pure states and Hamiltonians $H$, $||H||_{\infty}=1$. See main text for details on the sampling. The vertical lines in the box plot denote quartiles and functionals of interquartile ranges, while the mass of points starting around $.4$ denote outliers. The red line demarcates the bound defined in the main text.}
    \label{fig:histogram}
\end{figure}

\section{Entanglement Speed Limits}\label{sec2}

As noted above, the RT conjecture concerns bulk geometries culled from a renormalization flow; it states that geodesics in the bulk theory directly relate to the boundary entanglement \cite{Nozaki}. 
Oftentimes, a unitary flow (e.g., Hamiltonian evolution) is parameterized as a function of RG time. 
A natural route towards bounding entanglement entropy in the spirit of RT is therefore to bound the entropic speed limit. This prescription is central to proving the stability of the area law for one-dimensional systems \cite{PhysRevLett.111.170501}, as well as the logarithmic light-cone and entanglement spread in MBL systems \cite{kim2014local}. \par 
The small incremental entangling theorem (SIE), first posited by Kitaev \cite{bravyi2007upper}, gives an upper bound on the entanglement entropy rate of a pure state under unitary evolution: \begin{align}
    |\dot{E}(\rho(t))| \le c||H(t)||_{\infty} \log d_{A} , \, c \in \mathcal{O}(1). 
\end{align}

Here $E(\rho(t)):= S(\rho_{A}(t))$ denotes the entanglement entropy, $||\cdot||_{\infty}$ the operator norm, and $d_{A}$ the Hilbert space dimension of subsystem $\mathcal{H}_{A}$. The operator $H(t)$ is the generator of unitary evolution. The most general form of the SIE theorem involves ancillary qudits \cite{bravyi2007upper,PhysRevLett.111.170501}, which we do not consider here.

Proving the general SIE theorem has a long history, starting with Bravyi's work \cite{bravyi2007upper} and culminating with a proven bound $c = 18$ \cite{PhysRevLett.111.170501}. Since then, several works established tighter bounds on $c$ for the ancilla-assisted case--as extensively detailed in Refs \cite{Vershynina_2019}--with numerical work suggesting an optimal value $c=2$ both with and without ancillas. In the no-ancilla case, Bravyi's original proof yielded a $c(d)$ bound with $c(d) \to 1  $ as $d\to \infty$ \cite{bravyi2007upper}. More recently, a bound was given for no-ancilla SIE with $c=4$, independent of dimension $d$ and valid for mixed states \cite{hutter2012almost}. We offer a short proof here for the no-ancilla SIE that easily achieves a bound $c=2$ valid for mixed states. In the mixed state case we presume a pre-defined subsystem for which we take the entanglement entropy. \par 

We now detail our proof for no-ancilla SIE.
Let $\rho$ be a normalized density matrix acting on a Hilbert space $\mathcal{H} \cong \mathcal{H}_{A}\otimes \mathcal{H}_{A^{C}}$.  We take $H$ to be time-independent without loss of generality, and assume $d_{A}\le d_{A^{C}}$, where $d_{A}$ denotes the Hilbert space dimension. The entanglement rate is quickly derived as \begin{align}\label{eq:rate}
    |\dot{E}(\rho(t))| =|\Tr \rho(t) [H,\log \rho_{A}(t)\otimes 1_{A^{C}}]|.
\end{align} The Robertson-Schrodinger uncertainty relation then implies \begin{align}\label{eq:rs}
    |\Tr \rho(t)[H,Y(t)]| \le 2\sigma_{H}(t)\sigma_{Y}(t),
\end{align} where $Y(t) := \log \rho_{A}(t) \otimes 1 _{A^{C}}$. Here $\sigma_{X}^{2} = \Tr \rho(t) X^{2} - (\Tr \rho(t) X)^{2} \le ||X||_{\infty}^{2}$ for Hermitian $X$, with $||\cdot||_{\infty}$ denoting the operator norm. The variance \begin{align}
    \sigma_{Y}^{2}(t) = \Tr_{A} \rho_{A}(t)  \log^{2} \rho_{A}(t) - (\Tr_{A} \rho_{A}(t) \log \rho_{A}(t))^{2}
\end{align} is also known as the information fluctuation complexity (IFC) for the probability distribution $\bm{p}$, where $\bm{p}$ is the spectrum of $\rho_{A}$ \cite{bates}. We assume $\bm{p}$ has $n \le d_{A}$ non-zero components, where $d_{A}$ denotes the Hilbert space dimension of $A$. Then $
    \sigma_{Y}^{2}  \le \sum_{i=1}^{n}p_{i}\log^{2}p_{i} = M^{(2)}_{n} $, the second moment of the self-information. To bound the IFC, $\sigma_{Y}$, we follow and considerably simplify an approach first given in Ref. \cite{entropy}. Using the normalization condition $\sum_{i}p_{i}=1$, we define $p_{n} = 1-\sum_{i=1}^{n-1}p_{i}$. Simple calculus yields \begin{align}\label{eq:der}
    \frac{\partial M_{n}^{(2)}}{\partial p_{i}}=\log^{2} p_{i} + 2\log p_{i} - \log^{2}p_{n} - 2\log p_{n}  \\
    = (a_{i}-a_{n})(a_{i} + a_{n} -2),
\end{align} where we have defined $a_{i}= -\log p_{i}$. Clearly the maxima of $M_{n}^{(2)}$ vanish for each of the $n-1$ instances of Eq. \ref{eq:der}. The solutions $a_{i} = b$ for all $i$ implies a trivial solution $\bm{p} = (1/n,\ldots ,1/n)$, the uniform distribution. In the case $n = 2$, we find $p_{1} =  \frac{1}{2}\left(1+ \sqrt{1-4e^{-2}}\right) $ is a maxima, with the trivial solution as a minima. In this case one can easily check $\log 2$ is an upper bound for $M_{n}^{(2)}$.\par 
Our goal is to now show the trivial solution is the only maxima for $M_{n}^{(2)}$ when $n\ge 3$. We regard the set of variables $p_{1}, p_{2},r$ with $p_{n}=1-p_{1}-p_{2}-r$, $r\in [0,1)$. Without loss of generality, there are two cases to consider for a non-trivial solution: \begin{flalign}
    \nonumber \text{Case 1: } & p_{1} =p_{n}, \,  p_{2}p_{n}=e^{-2}, \\ \nonumber
    \nonumber \text{Case 2: }  & p_{1}p_{n}= e^{-2} , \, p_{2}p_{n} = e^{-2}.
\end{flalign} We utilize Gr\"{o}bner bases to write Case 1 as the null space of the following set of equations: $\{p_{n}-p_{1},  p_{2} +r+ 2p_{n}-1, 2p_{n}^{2}+(r-1)p_{n}+e^{-2}\}$. A check of the discriminant of the last polynomial yields $(1-r)^{2}-8e^{-2} < 0$ for any $r$. For Case 2 a Gr\"{o}bner basis is $\{2p_{1}+r+p_{n}-1,2p_{n}+r+p_{n}-1,p_{n}^{2}+(r-1)p_{n}+2e^{-2}\}$, with the same discriminant as before. Therefore, the uniform distribution is the only maxima of $M_{n}^{(2)}$ for $n \ge 3$. For $n\ge 2$ we then have $\sigma_{Y}^{2} \le \log^{2} n$ and a final bound \begin{align}\label{eq:sie}
        |\dot{E}(\rho(t))|\le 2||H||_{\infty}\log d_{A}.
    \end{align}
While the logarithmic dependence of the above bound can be saturated \cite{bravyi2007upper}, most states have a slow entanglement spread under the unitary evolution induced a large plurality of typical Hamiltonians \cite{hutter2012almost}. In Fig. \ref{fig:histogram}, we depict a histogram of two million realizations of two-qubit Hamiltonians and pure states. The Hamiltonians are generated by uniformly sampling eigenvalues from $[-1,1]$ (thus ensuring $||H||_{\infty} =1$), and then applying a Haar unitary. The pure states were generated by uniformly sampling a full rank Schmidt vector $\bm{p} = (p_{1},1-p_{1})$ (the Haar unitary is then interpreted as in the Schmidt basis of $\rho$). The quartiles compared to the outlier spread gives a clear indication that most quantum states will not saturate the SIE bound.

\section{Connecting to Quantum Information  Geometry}\label{sec3}
We now use the SIE theorem as a bridge to consider bulk geometry by identifying computable bulk quantities and metrics related to the flow entanglement entropy. The RT theorem indicates that minimal cuts in the bulk should be proportional to the boundary entanglement; thus, making an identification with the flow entanglement entropy helps preserve this correspondence. As we show, these bulk metrics comprise two components: a quantity related to distances in the projective Hilbert space, and the information fluctuation complexity (IFC), described above. Approximations to either of these quantities yields a variety of bulk metrics: we find that there is a trade-off between approximations which hew most closely to the entanglement flow, and those which are easier to compute.\par 

\par 
We start by examining approximations to the two quantities on the RHS of Eq. \ref{eq:rs}.
In practice, $||H||_{\infty}$ is expensive to calculate compared to the Frobenius norm $||H||_{F}:= \sqrt{\Tr H^{\dagger}H}$. Moreover, we have $||A||_{\infty} \le ||A||_{F}\le \sqrt{r}||A||_{\infty}$ for any rank $r$ operator $A$; thus, we can approximately bound $\sigma_{H}$ by $||H||_{F}/\sqrt{d}$, where $d:= d_{A}d_{A^{C}}$. For traceless $H$ the factor $||H||_{F}/\sqrt{d}$ corresponds to the infinitesimal root-mean Fubini-Study arc-length, where the mean is taken over an orthonormal set of states. \par  Returning to the IFC, we decompose $\sigma_{Y}(t)$ into a running average and fluctuating component (with respect to $t$), $\sigma_{Y}(t) = \bar{\sigma}_{Y} + \tilde{\sigma}_{Y}(t)$. We denote $\braket{\cdot}$ as averaging over the set of flow states. Thus we have \begin{align}
    E(\rho(t)) \le  2\left(\bar{\sigma}_{Y}\int_{0}^{t} dt' \sigma_{H}(t') + \int_{0}^{t} dt' \tilde{\sigma}_{Y}(t') \sigma_{H}(t')\right).\end{align}
    Making the replacement $\sigma_{H} \to ||H||_{F}/\sqrt{d}$ and assuming $\braket{\tilde{\sigma}_{Y}(t)}  \approx 0$, we arrive at 
    \begin{align}
    \braket{E(\rho(t))}\lesssim  \frac{2 \braket{\bar{\sigma}_{Y}}}{\sqrt{d}}\int_{0}^{t} dt' ||H(t')||_{F}= 2\braket{\bar{\sigma}_{Y}}D_{X},
\end{align} 
where we have defined \cite{yu2019bulk} \begin{align}\label{eq:Dx}
    D_{X}(t):= \frac{1}{\sqrt{d}}\int_{0}^{t} dt' ||H(t')||_{F}.
\end{align}
We interpret $D_{X}$, first suggested as a bulk metric in the context of the Wegner Wilson Flow in Ref. \cite{yu2019bulk}, as quantifying the average strength of rotation induced by the unitary process, integrated over the flow. The term $D_{X}$ is, for traceless $H$, the arc length of the unitary $U$ under a standard bi-invariant Riemannian metric \cite{Dowling_Nielsen_2006}. Different metric choices that penalize many-qubit Hamiltonian operators give rise to a geometrical notion of quantum circuit complexity \cite{Dowling_Nielsen_2006,PhysRevLett.120.121602}. We also define an intermediate bulk quantity, $D_{XY}:= \int ||H||_{F}\sigma_{Y}/\sqrt{d}$, that incorporates both the root-mean Fubini Study metric and the IFC. \par 

In addition to defining a bulk metric in terms of $\sigma_H$, we can also relate $\sigma_H$ to the efficiency of the unitary flow. Note that for pure states $\rho = \ket{\psi}\bra{\psi}$, the term $\sigma_{H}^{2} \equiv g_{tt}$ is simply the diagonal part of the Fubini-Study metric, $g_{\mu\nu} = \Re(Q_{\mu\nu})$ under unitary evolution $U(t)$ where $Q_{\mu\nu}$ is the
quantum geometric tensor defined as \begin{equation}\label{eq:qgt}
    Q_{\mu\nu} = \bra{\partial_{\mu}\psi}\left(1 - \ket{\psi}\bra{\psi}\right)\ket{\partial_{\nu}\psi},
\end{equation}
The imaginary part of $Q_{\mu\nu}$ is then proportional to the Berry curvature \cite{Neupert2013}. 
An arc-length in the projective Hilbert space determined by a continuously-parameterized unitary is defined (up to a universal constant) by $d_{FS}(\psi(t)):=\int_{0}^{t} dt'  \sqrt{g_{tt}}=\int_{0}^{t} dt' \sigma_{H}(t')$. The efficiency of $U(t)$ with respect to flow state $\psi(t)$ is characterized by the ratio of the geodesic arc-length to the arc-length taken by $U$ \cite{cafaro2020geometric}: \begin{equation}\label{eq:eff}
    \varepsilon(\psi(t)) := \frac{\cos^{-1}|\braket{\psi(t)|\psi(0)|}}{d_{FS}(\psi(t))}=\frac{\cos^{-1}|\braket{\psi(t)|\psi(0)}|}{\int_{0}^{t}dt' \sigma_{H}(t')}.
\end{equation} Using Eq. \ref{eq:rs} and the Fubini-Study definition of $\sigma_{H}$, we extract an upper bound on the entanglement entropy of the boundary state $\psi(t)$ via \begin{align}\label{eff_bound}
    E(\rho(t)) \le 2\cos^{-1}\left(|\braket{\psi(t)|\psi(0)}|\right)\varepsilon(\psi(t))^{-1}\log d_{A}.
\end{align} That the entanglement entropy bound scales inversely with the efficiency is consistent with shallow-depth local quantum circuits obeying area-law entanglement.

Thus we see that our entanglement entropy bound comprises two components: a distance measure in the projective Hilbert space, and a complexity measure (IFC) particular to the bipartition. While the complexity measure is difficult to numerically obtain (we effectively need to measure the entanglement entropy at each flow step), the divergence between entanglement growth and unitary rotation becomes manifest, as we now explore in the context of MBL.

\begin{figure}
    \centering
    \includegraphics[trim = 1cm 0cm 2.5cm 0cm, clip,width=\linewidth]{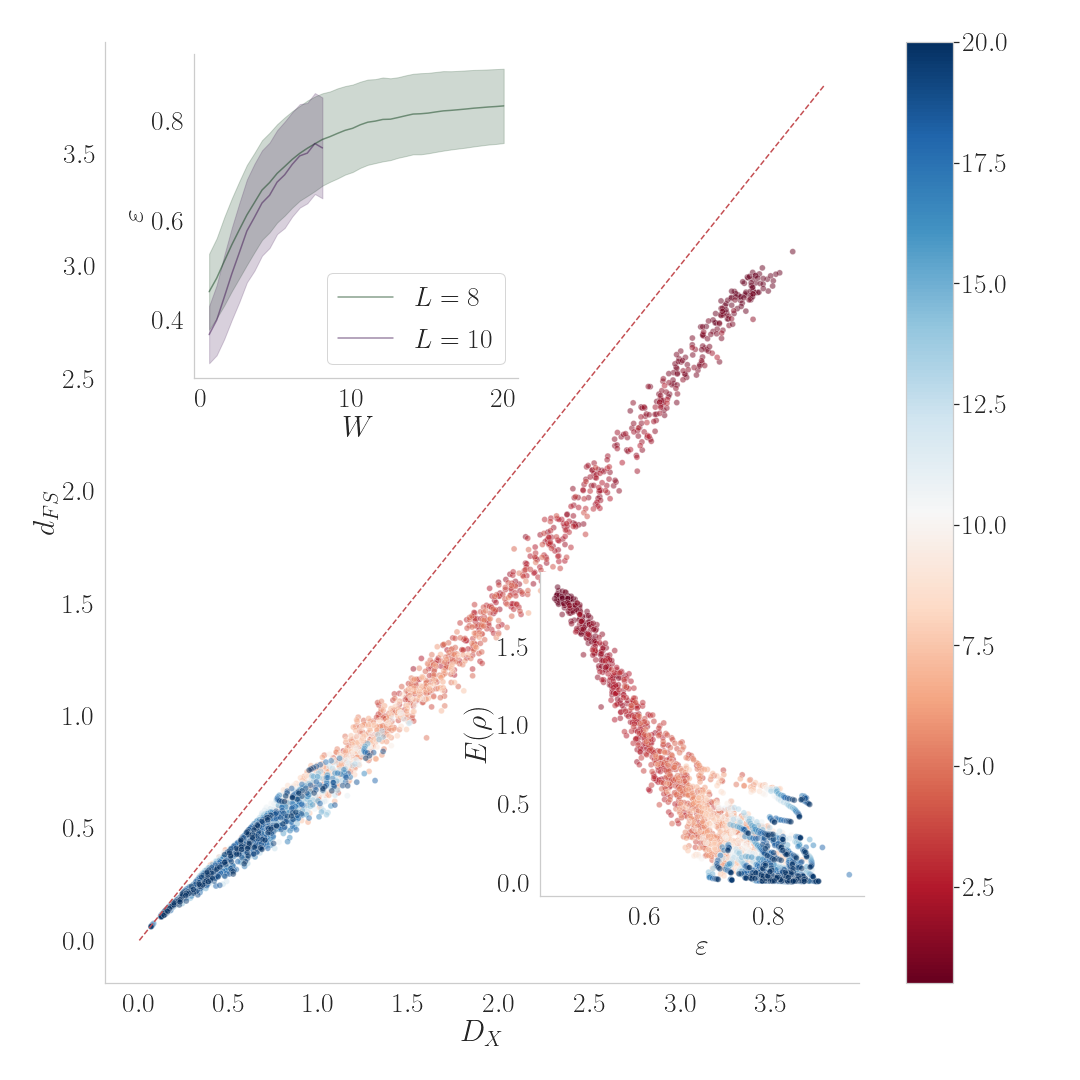}
    \caption{Average WWF arc-length $d_{FS}$ against the root-mean WWF arc-length $D_{X}$ for $L=8$. Colors denote varying disorder strengths; the red diagonal line is a guide to the eye. (\textit{Left inset}) Median efficiency against disorder strength $W$ for both $L=8$ and $L=10$. Shading denotes standard deviation, estimated via bootstrap sampling. (\textit{Right inset}) Average entanglement entropy against WWF efficiency for $L=8$, with color coding the same as the main plot.}
    \label{fig:eff}
\end{figure}

\begin{figure}
    \centering
    \includegraphics[width=\linewidth,trim = 0cm 0cm 0cm 0cm ,clip]{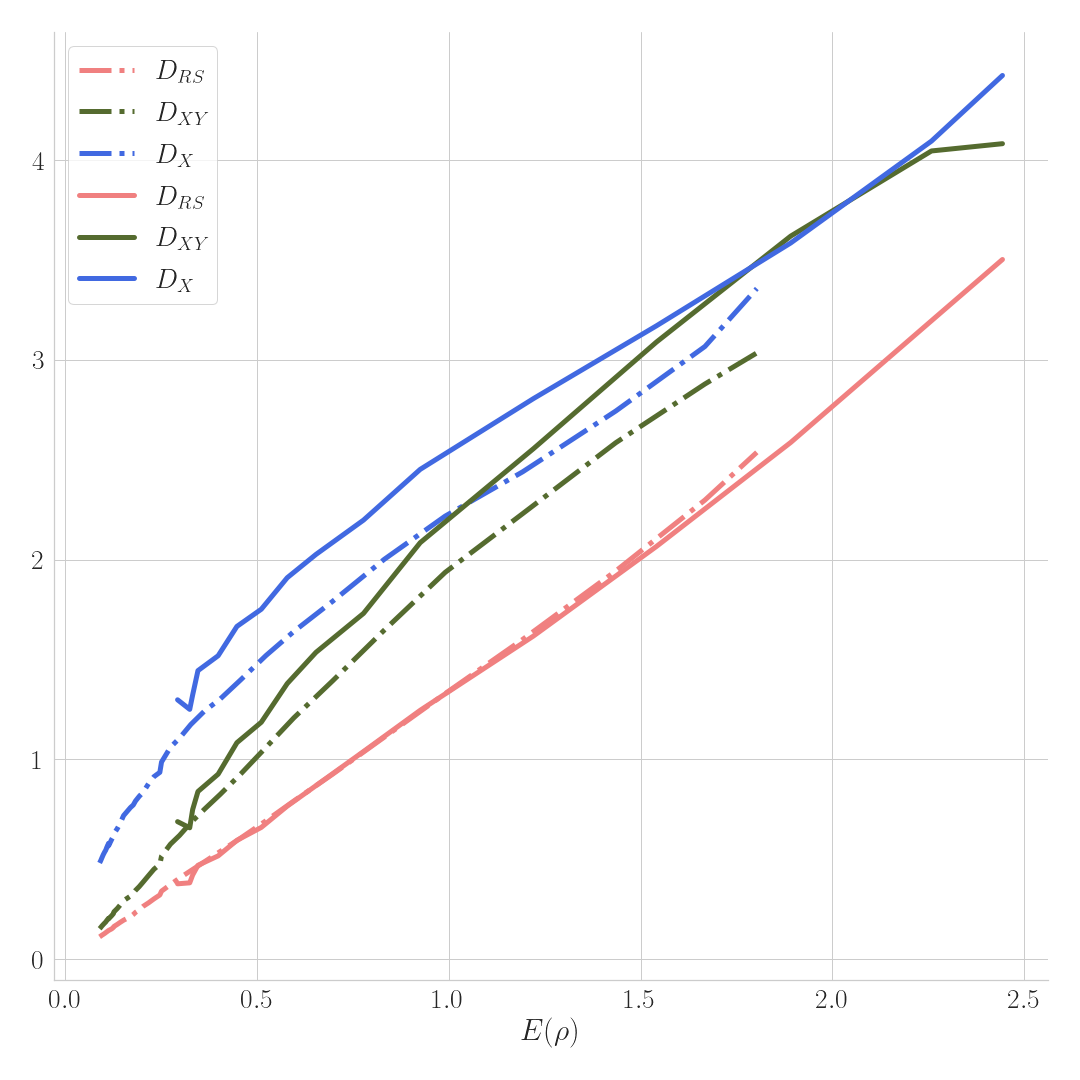}
    \caption{Three bulk entanglement measures against $E(\rho)$. $D_{RS}$ denotes (half) the integrated RHS of Eq. \ref{eq:rs}, while $D_{XY}:= \int ||\eta||_{F}\sigma_{Y}/\sqrt{d}$. The dashed lines denote $L=8$, while the solid lines denote $L=10$. We perform averaging within a disorder realization before averaging over disorder realizations.}
    \label{fig:bounds}
\end{figure}

\section{Wegner Wilson Flow and MBL}\label{sec4}

The entanglement analysis given above is applicable to any continuous unitary process; however, our interests lie in one unitary particularly relevant to MBL and generating $\ell$-bits: the Wegner Wilson flow (WWF). The WWF is fundamentally a diagonalization protocol that continuously evolves an orthonormal set of initial states into eigenstates of a given Hamiltonian. For Hamiltonian $H_{0}$ and initial basis $\{\psi_{0}\}$, the WWF is succinctly expressed by the differential equation \begin{align}\label{eq:unit}
     \frac{dU(\beta)}{d\beta} = \eta(\beta)U(\beta),
\end{align} whereby we define $H(\beta) = U(\beta)H_{0}U^{\dagger}(\beta)$ as the flow Hamiltonian. The term $\eta(\beta) = [H_{d}(\beta),H(\beta)]$ is the commutator of the diagonal component (with respect to the $\psi_{0}$ basis) of the flow Hamiltonian, $H_{d}(\beta)$, with $H(\beta)$. We define the variance of the WWF as \begin{nalign}\label{eq:wwfvar} V(\beta) = \langle V_{0}(\beta) \rangle_{0} = \braket{(H(\beta)-H_{d}(\beta))^{2}}_{0},\end{nalign} where $\braket{\cdot}_{0}$ denotes averaging over $\{\psi_{0}\}$, and $V_{0}(\beta) = \braket{\psi_{0}|(H(\beta)-H_{d}(\beta))^{2}|\psi_{0}}$. The WWF ensures $V(\beta) \to 0$ monotonically; in particular, the decay of off-diagonal elements of $H(\beta)$ scales exponentially with $[\beta] = \text{energy}^{-2}$ \cite{itto2012geodesic,monthus2016flow}. Moreover, the WWF induces a geodesic flow with respect to the Fubini-Study metric given appropriate constraints on $H_{0}$ and $\{\psi_{0}\}$ \cite{itto2012geodesic}. Numerical studies have demonstrated the WWF generally produces more quasilocal (in the sense of support on the lattice) integrals of motion in the MBL phase than other flow equation methodologies   \cite{savitz2017stable,PhysRevB.100.115136}. The WWF is also easier to implement, though it is susceptible to numerical stiffness issues \cite{savitz2017stable}.\par 
 As a testbed for our analytics, we consider the prototypical MBL Hamiltonian, the one-dimensional spin-$1/2$ Heisenberg model with on-site disorder and open boundary conditions. The Hamiltonian is given by \begin{equation}\label{eq:ham}
    H = \frac{1}{4}\sum_{i}\bm{\sigma}_{i}\cdot \bm{\sigma}_{i+1} + \frac{h_{i}}{2}\sigma_{i}^{z},
\end{equation} where $h_{i} \in [-W,W]$ is sampled uniformly. A large body of numerical and theoretical work has placed the MBL transition in the thermodynamic limit at $W_{c} \approx 3.8$ for this model, though the dependence on system size has recently come into question \cite{pal2010,luitz2015many,vidmar2020}. We perform WWF within the zero magnetization sector, and work within the computational $\sigma_{z}$ basis. We utilize the same numerical procedure and tests for convergence outlined in previous works \cite{yu2019bulk,PhysRevB.100.115136}. We performed our analysis on $L=8$ and $L=10$ system sizes, and average within a disorder realization before averaging over realizations. Due to calculating the IFC at each stage of the flow, we are relegated to smaller system sizes. For the case of $L=10$, we choose to calculate the IFC at each flow step for only a subset of flow states. However, the qualitative results of our analytics given above is still clear. \par

Fig. \ref{fig:eff} depicts the median WWF efficiency (as defined in Eq. \ref{eq:eff}) as a function of disorder, where we see a clear monotonic climb towards unity upon entering the MBL phase. This monotonicity is coincident with the increasing quasilocality of the integrals of motion: in the infinite disorder limit, the eigenstates are effectively product states, and therefore the flow is essentially geodesic with respect to the Fubini-Study metric. Fig. \ref{fig:eff} also shows $d_{FS}$, the average WWF arc-length from initial to final state, against $D_{X}$, the root-mean WWF Fubini-Study arc-length. By Jensen's inequality, $D_{X}$ is larger than $d_{FS}$, though this inequality becomes parametrically weaker upon entering the MBL phase. Fig. \ref{fig:eff} further depicts the average entanglement entropy of the eigenstates against the WWF efficiency. Consistent with our claim above, a higher efficiency correlates with a lower entanglement entropy.
\par

\begin{figure}[ht]
    \centering
    \includegraphics[width=\linewidth,trim=.8cm 1cm 1cm 1cm,clip]{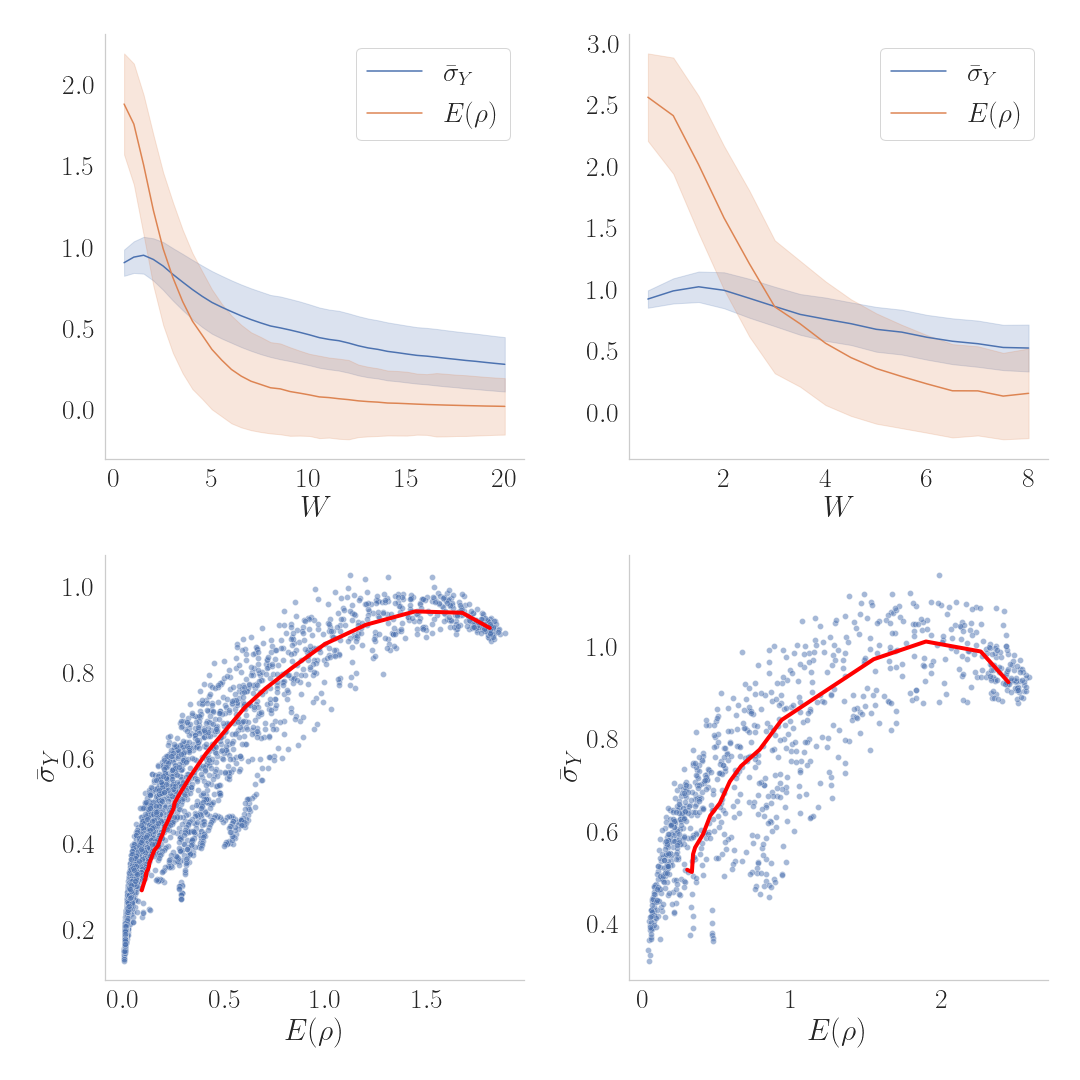}
    \caption{(\textit{Top}) Average (over flow states) IFC $\bar{\sigma}_{Y}$ and eigenstate entanglement entropy against disorder strength $W$ for $L=8$ (left) and $L=10$ (right). Shading denotes standard deviation. (\textit{Bottom}) Average IFC against $E(\rho)$ for $L=8$ (left) and $L=10$ (right). The points represents values averaged within a disorder realization, while the red line denotes trend after averaging over disorder realizations.}
    \label{fig:running_variance}
\end{figure}

The Fubini-Study metric is intimately tied to the diagonalization rate of the WWF. In particular, we have \begin{align}
    g_{\beta\beta}(\psi) = -\frac{1}{2}\frac{dV_{0}(\beta)}{d\beta},
\end{align} where $\psi$ is the flow state for initial state $\psi_{0}$ \cite{yu2019bulk}. As the disorder terms $h_{i}\sigma_{i}^{z}$ are diagonal in the spin configuration basis, and $\lim_{\beta\to\infty}V_{0}(\beta) = 0$, the energy functional $\int g_{\beta\beta}d\beta \propto V_{0}(0)$ is independent of disorder strength or realization for every flow state $\psi$. 
Our recasting of the Fubini-Study metric then implies \cite{monthus2016flow,yu2019bulk}\begin{align}
        D_{X} = \int d\beta \left(-\frac{1}{2d}\frac{dV(\beta)}{d\beta}\right)^{1/2}.
    \end{align}
    In Fig. \ref{fig:bounds}, we depict $D_{X}$ against the average eigenstate entanglement entropy $E(\rho)$ (we suppress the averaging notation for brevity), as well as  $D_{RS}:=\int  \sigma_{\eta}\sigma_{Y} $, proportional to the integral of Eq. \ref{eq:rs}. We note that $D_{RS}$ is strongly linear throughout the phase diagram, in strong contrast to $D_{X}$. The nonlinearity in $D_{X}$ is most manifest for small $E(\rho)$ (large disorder strength $W$), as in the infinite disorder limit both $E(\rho)$ and $D_{X}$ should approach zero. 
    
    The central difference between $D_{X}$ and $D_{RS}$ lies in $\sigma_{Y}$, the information fluctuation complexity (IFC). In Fig. \ref{fig:running_variance}, we depict the average IFC $\bar{\sigma}_{Y}$ (recall this average is over the flow trajectory, then over the set of eigenstates) against disorder strength $W$. We simultaneously plot $E(\rho)$ against $W$, and note that $E(\rho)$ decreases at a faster rate than $\bar{\sigma}_{Y}$. The lower panels of Fig. \ref{fig:running_variance}, wherein we plot $\bar{\sigma}_{Y}$ against $E(\rho)$, more clearly shows the discrepancy. \par 
    The consequence of this nonlinearity is that the degree of rotation quantified by $D_{X}$ fails to linearly correlate with the degree of entanglement growth as the system enters the MBL phase. To phrase this differently, a measure of unitary rotation is inequivalent to a measure of entanglement growth (e.g., a unitary that is a tensor product of unitaries acting on each side of a bipartition does not generate entanglement over that bipartition). \par

    It is interesting to note that, in the context of cMERAs, the Fubini-Study/Bures metric serves as a suitable bulk measure for the entanglement entropy, unlike the nonlinear relation presented in Fig. \ref{fig:bounds}. We suspect this difference stems from the cMERA hyperbolic geometry in the thermodynamical limit, in which geodesics (minimal surfaces) are lines extending into the bulk, and the Fubini-Study metric effectively measures the strength of disentanglers \cite{Nozaki,swingle2012entanglement,qi2013exact}. 
    In the present WWF context, the unitary bulk represents an \textit{energy diagonalization} flow, as reflected in the energy functional $V_{0}(0) = \int d\beta g_{\beta\beta}$; the goal of the WWF is a projective Hilbert space trajectory such that $V(\beta)$ optimally decays. As demonstrated here, that goal is generally at odds with a flow that optimally generates entanglement. We leave open the question of whether a more natural bulk metrical construction quantifies the entanglement entropy generated by the WWF.\par

    To briefly conclude, we probed the bulk geometry of the WWF using measures of both efficiency and entanglement growth. We gave a new proof of the small incremental entangling conjecture, and related the resulting bounds to bulk geometry and computational metrics. We used the WWF in the context of MBL as a numerical testbed, and assessed the WWF efficiency as a function of disorder strength. We demonstrated where bulk metrics failed to cleanly relate to the boundary entanglement entropy, and identified the information fluctuation complexity as a key quantity separating quantum evolution measures from entanglement measures.

\section*{Acknowledgments}
GH acknowledges useful conversations with Benjamin Villalonga, Di Luo, Abid Khan, Ryan Levy, and Eli Chertkov. This work made use of the Illinois Campus Cluster, a computing resource that is operated by the Illinois Campus Cluster Program (ICCP) in conjunction with the National Center for Supercomputing Applications (NCSA) and which is supported by funds from the University of Illinois at Urbana-Champaign.  We acknowledge support from the Department of Energy grant DOE DESC0020165.

\bibliography{library}
\end{document}